\documentclass[twocolumn,secnumarabic,amssymb, nobibnotes, aps, prb]{revtex4-1}

\usepackage{color}
\usepackage{pstricks}
\usepackage{graphicx}
\usepackage{amsmath}

\usepackage{graphicx}
\newcommand\scalemath[2]{\scalebox{#1}{\mbox{\ensuremath{\displaystyle #2}}}}

\newcommand{\beq}{\begin{equation}}
\newcommand{\eeq}{\end{equation}}

\newcommand{\beqa}{\begin{eqnarray}}
\newcommand{\eeqa}{\end{eqnarray}}

\newcommand{\ora}{\overrightarrow}

\usepackage{slashed}
\let\beq=\begin

\let\a=\alpha
\def\be{\begin{equation}}
\def\ee{\end{equation}}
\def\bea{\begin{align}}
\def\ea{\end{align}}
\let\s=\slashed
\def\eps=\epsilon
\begin{document}

\title{Chern-Simons terms in the 3D Weyl semi-metals}
\author{Hamid Omid}
\affiliation{Department of Physics and
Astronomy, University of British Columbia\\6224 Agricultural Road,
Vancouver, British Columbia V6T 1Z1}
\date{May 5, 2014}
\maketitle

\section{ABSTRACT}
Based on some theoretical arguments, it has been suggested that electromagnetic response of 3D Weyl semi-metals  with non-zero chiral- chemical potential may have a Chern-Simons term, $\frac{1}{2}k_\mu \epsilon^{\mu\nu\rho\sigma}F_{\nu\rho}A_\sigma$, in their effective action for the gauge field. An independent numerical study has shown that such a term is absent in a similar system. In this paper, we investigate the non-equilibrium and equilibrium response of 3D Weyl semi-metals. We argue that the controversy in literature stems from the difference in response of these two distinct states. We then develop a method to deal with well-known ambiguities in quantum electrodynamics in $3$D~(QED$_{3+1}$) with non-zero chiral-chemical potential and calculate the Chern-Simons term unambiguously. We find that  time-like Chern-Simons term can exist in non-equilibrium conditions. We observe that there does not exist any chiral-magnetic effect  in equilibrium and anomalous Hall effect replaces it.
\section{INTRODUCTION}
Recently, 3D Weyl semi-metals have gained much attention. 2D Weyl semi-metals became of interest by experimental fabrication of graphene\cite{Geim} , although they were expected to have novel properties for a while \cite{Gordon}. Weyl semi-metal is a phase of matter in which the valence band touches the conduction band at certain points and the dispersion around the  so called 'Weyl Point' takes the form of relativistic dispersion. Although 2D Weyl semi-metal phase is sensitive to the perturbations and can be gapped easily by breaking of underlying symmetries, 3D Weyl semi-metals show more robust behaviour\cite{Herring,Murakami}. In the absence of $P$ or $T$ it can be shown that there is a finite region of internal parameters that leads to a gap-less state\cite{Murakami}. This can be understood by noting that Weyl nodes are associated with a pseudo-charge which are conserved. Conservation of this charge makes the gaping hard. In order to gap the spectrum, Weyl nodes that are located at different momentums need to meet or interact with each other. 
Based on the above consideration, there are a few suggestion for realizing 3D Weyl semi-metals experimentally \cite{Wan,Burkov1}. 
\\

It has been theoretically suggested that presence  of non-zero chiral- chemical potential in effective Hamiltonian of Weyl semi-metals, which violates emergent Lorentz symmetry,  can result in induced Chern-Simons term in photon effective Lagrangian,
\begin{equation}
\mathcal{L}_{\text{CS}}=\frac{1}{2}k_\mu \epsilon^{\mu\nu\rho\sigma}F_{\nu\rho}A_\sigma
\end{equation}
in which $k_\mu$ is the  Chern-Simons coefficient,  $\epsilon^{\mu\nu\rho\sigma}$ is the anti-symmetric tensor in 3D and $F_{\nu\rho}$ is the field strength of vector potential $A_\sigma$.
Chern-Simons term has the fascinating feature that in the absence of electric field there would be an electric current solely induced by a magnetic field. This feature, called 'chiral-magnetic effect', is evident in associated current derived from gauge invariance ,
\begin{align}
&\rho={\bf k}.{\bf B} \nonumber \\
&{\bf j}={\bf k}\times{\bf E}-k_0{\bf B}
\end{align}

Presence of Chern-Simons term in Lorentz violating QED$_{3+1}$, the effective theory describing Weyl semi-metals, is a well-known feature \cite{Perez,Jackiw1,Jackiw2}. It is believed though that $k_\mu$ is ambiguous and depends on the regulator used to regulate linear divergences of theory. 
It is suggested \cite{Marcel} that if the correct regulator gets used in calculations, chiral-magnetic effect would vanish. 
In this paper, we discuss a possible explanation for the discrepancy between different results. We will argue that the discrepancy stems from different linear responses associated with a system. We then try to find a proper regulator and see whether such a regulator forbids presence of chiral-magnetic effect.

\section{THE MODEL AND REGULATOR}\label{model}
\begin{figure}[ht]
\begin{center}
\includegraphics[scale=.45]{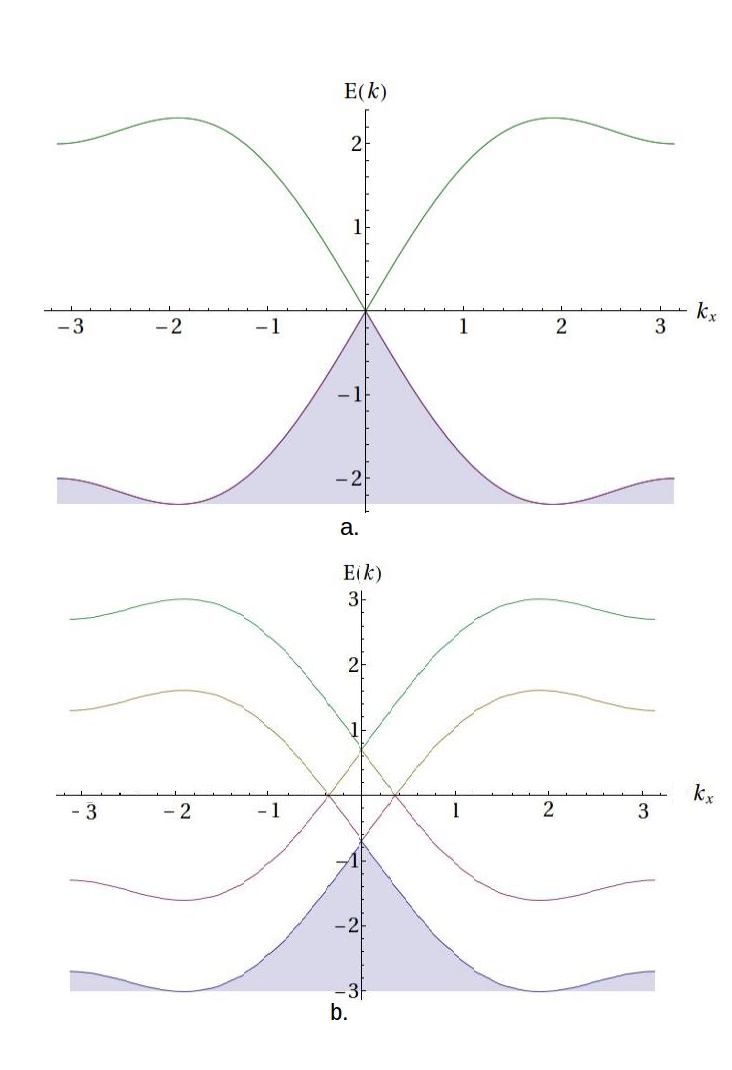}
\end{center}
\caption{{\footnotesize The band structure of our model for $\epsilon=6t$ in units of $\lambda$ sketched at $k_y=k_z=0$ a. Doubly degenerate Dirac point for b=(0,0,0,0) b. Shifted Dirac points for b=(0,0.7,0,0)  .}}
\label{well1}
\end{figure}

\
In the rest of the paper, we investigate the standard model describing a 3D Topological Insulator \cite{Qi83,Fu105}, defined by the momentum space Hamiltonian
\begin{align}\label{H_0}
H_0({\bf k})=&2\lambda \sigma_z(s_x sin(k_y)-s_y sin(k_x))+2\lambda_z\sigma_y sin(k_z)\nonumber \\&+\sigma_x M({\bf k})
\end{align}
with $\sigma$ and $s$ the Pauli matrices acting in orbital and spin space respectively and $M({\bf k})=\epsilon-2t\sum_i cos(k_i)$.  Without loss of generality we restrict our model to the case that $\lambda_z=\lambda$. 
We constrain our parameters so that the system lives in trivial phase. This can be done by implementing $\epsilon=6t$, resulting in $M({\bf k}=\Gamma)=0$ such that $\Gamma=0$ . Although this Hamiltonian realizes Weyl semi-metal phase, this phase only exist at a single point in parameter space given by  $\epsilon=6t$.  To realize the Weyl semi-metal phase that is stable, we add the following term, $H_1$, to our original Hamiltonian,
\begin{equation}
H_1({\bf k})=b_0 \sigma_y s_z+{\bf b}.(-\sigma_x s_x,\sigma_x s_y,s_z)
\end{equation}
$b_0$ and ${\bf b}$ terms in perturbation break $\mathcal{P}$ and $\mathcal{T}$ respectively. Violation of  $\mathcal{P}$ or $\mathcal{T}$ is needed to realize Weyl semi-metal phase in a finite region of phase space\cite{Herring,Murakami}. As can be expected from symmetry effect of this term, it has magnetic origin. It can be introduced by magnetic doping of a Weyl semi-metal system and has been observed experimentally to be present in topological insulators\cite{Grushin,Chen}.\\
The low energy limit of $H_0$ can be written in familiar form of Dirac Hamiltonian,
\begin{equation}\label{H_02}
H_0(k)=2\lambda{\bf \a. k}+\beta m
\end{equation}
with $\alpha$ defined in [\ref{A1}] and in special case of  $\epsilon=6t$, $m=0$ . In [\ref{A1}],  $H_1$ and second order contribution of $H_0$ are written in the same basis, here we recall the results,
\begin{equation}
H_1=b_0 \gamma_5+{\bf \a.b}  \gamma_5
\end{equation}
 There are two different linear-responses that can be obtained from this theory. These two responses have been explored in \cite{Marcel,Burkov2}, for example. After adding $H_1$ to the original Hamiltonian the place of Weyl nodes in momentum space changes. As a result, the electrons which have been in the ground state previously need to eventually move to a new region of phase space  FIG.~\ref{well1}. In other words, system will form a metastable state that will eventually decay into the actual ground state. Either $H_1$ can be added to the Hamiltonian as a perturbation and the conductivity  of the systems can be studied before decaying into actual ground state or the conductivity  can be studied in equilibrium state.  The two responses are quite different and will result in different effective actions for a coupled gauge field. \\

In section \ref{derivation}, we study the former case in which we assume the system stays in meta-stable ground state. In this case the low-energy theory is given by the Weyl nodes around ${\bf k}=0$. Fortunately, the latter case does not need an independent calculation and can be understood without use of any regulator. As explained, the Weyl nodes move in momentum space and we need to linearize the Hamiltonian around that points. We can perform the same procedure in [A] and find out that the effective Hamiltonian would be given by,

\begin{equation}
H_0(k)=2\lambda{\bf \a. (k-k_{+}})+2\lambda{\bf \a. (k-k_{-}})
\end{equation}
in which $k_{+}$ and $k_{-}$ are the location of the new Weyl points. The above Hamiltonian is the same as the Hamiltonian that is studied in Balents, et al \cite{Burkov1}. In this case, the chiral-chemical potential is absent which means that the theory is well-defined and there is no dependence on the regulator. Balents, et al have shown that the conductivity gets the form of  an anomalous Hall effect.
This shows that the chiral-magnetic effect will be absent in equilibrium and only anomalous Hall effect survives.

\section{Regulation Problem}\label{divergences}
\
It is well-known that a finite quantum field theory~(QFT) with linear divergences may include ambiguities. By finite QFT that is linearly divergent, we mean a QFT that has a set of finite correlators which are superficially divergent. The presence of such ambiguities can be understood by considering elementary integrals that are linearly divergent. Consider a function $f(x)$ which is finite at infinity but non-zero. Let us consider the following integral,
\begin{equation}
\int_0^\infty dx~(f(x+a)-f(x))\sim a~f'(x)|_0^\infty+O(a^2)
\end{equation}
although the first term looks like the same as the second term with an inconsequential change of variables, the  integral depends on that change of variable through its dependence on $a$. In the presence of linear divergences, although the integrals may be finite, the way that they get manipulated changes the final result. For example, a change of variable before combining individual integrals may change the final result.\\
It is known \cite{Perez,Jackiw1,Jackiw2} that extended QED$_{3+1}$ with non-zero chemical potential, which breaks $\mathcal{PT}$ is a finite but lineally divergent QFT. As a result, some of the correlators, for example photon polarization tensor, depend on the way that different Feynman diagrams get combined together. 
To deal with this ambiguity, a regulator should be chosen so that all of the correlators become convergent and analytic. One can think of different regulators as different ways of combining the Feynman diagrams. 
The dependence of the results on the choice of regulator is a consequence of the importance of high-energy theory beyond its contribution to linearized action. To get an unambiguous result, one can return to the full theory that we have derived our effective theory from and perform the calculations using that theory. 
In this paper, we take advantage of having such a theory and find the appropriate regulator by analyzing it. In addition to finding the effective theory by finding the low-energy limit Hamiltonian, we find the next order terms which are higher in powers of momentum. This higher order term will regulate the linear divergence as it can be seen by power counting. Let us borrow the formula that we derive for two-point function in section \ref{derivation} and count the powers of momentum we are integrating, $q$.  
\begin{widetext}
\begin{align}
\Pi_{\mu \nu}(p)&=-\frac{8i\epsilon^{\mu\nu\rho\sigma}}{(2\pi)^4}\int d^4q~\frac{-2q.b q_\rho(q+p)_\sigma+q^2b_\rho(q+p)_\sigma+m^2(q-p)_\rho b_\sigma}{(q^2-m^2)^2((q+p)^2-m^2)}
\end{align}
\end{widetext}

This integral suffers from superficial linear divergence. We call it superficial as the final result for the integral is finite even without use of any regulator. It is clear that by adding a higher order term to the propagator denominator, $\frac{i}{\s{p}-m}$, the linear divergence gets removed. We use this trick to regulate our integrals.

\section{DERIVATION OF CHERN-SIMONS TERM}\label{derivation}
\
In this section, we derive the Chern-Simons coefficient using our physical regulator. In section \ref{model}, we discussed the rise of QED$_{3+1}$ as the effective theory describing the quasi-particles in Topological Insulator heterostructure. The Lagrangian for QED$_{3+1}$ is given by
\begin{equation}
\mathcal{L}_{\text{QED}}=\overline{\psi}(i\s{\partial}-m- \s{A})\psi
\end{equation}
$\psi$ is the fermionic field, $A$ is the gauge field which is coupled to $\psi$ by minimal coupling resulting from Peierls substitution in equation \ref{H_0} and $m$ is the fermionic mass which we keep to regulate IR divergences and put zero at the end. 
We showed that presence of non-zero $b$ results in a term that breaks $\mathcal{PT}$. As it is shown in [\ref{A1}], this term gets the form of chiral-chemical potential. The Lagrangian for this potential is given by,
\begin{equation}
\mathcal{L}_b=-\overline{\psi}\gamma_5\s{b}\psi
\end{equation}
As the last piece, we have another term in our Lagrangian that plays the role of regulator. In [\ref{A1}], we derived the regulator and showed that it can be written as a momentum dependent mass term and has the form of,
\begin{equation}
\mathcal{L}_{\text{regulator}}=\frac{t}{\lambda}\overline{\psi}{\ora{\partial}^2}\psi
\end{equation}\

\begin{figure}[ht]
\begin{center}
\includegraphics[scale=.23]{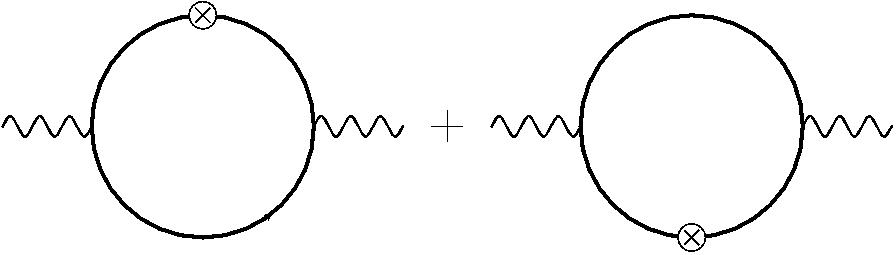}
\end{center}
\caption{{\footnotesize The vertex can be placed on both of the internal fermion lines resulting in two Feynman diagrams.}}
\label{oneloopvertex}
\end{figure}

We emphasize that this regulator breaks the Lorentz invariance manifestly, which ensures that our final result for Chern-Simons term is not an artifact of keeping Lorentz invariance in low energy theory.
The Chern-Simons's term can be calculated using the photon two-point function expansion around $p_\mu=0$. We treat the chiral-chemical potential as an extra vertex and find the two-point function order by order. The two point function can be found using usual Feynman rules. As is mentioned in \cite{Perez}, the 1-loop calculation results in exact form of induced Chern-Simons term and we don't need to go to higher orders to find further contributions. We use the Clifford Algebra between Dirac matrices and the identities follow from this fundamental anti-commutation relation. In particular, we use the fact that for $n~\epsilon \mathcal{N}$, $tr(\gamma_{\mu_1}...\gamma_{\mu_{2n+1}})=0$ and $tr(\gamma_{\mu}\gamma_{\nu}\gamma_{\rho}\gamma_{\sigma}\gamma_{5})=-4i\epsilon^{\mu\nu\rho\sigma}$.  Using these identities we find an integral form for photon two-point function,
\begin{widetext}
\begin{align}\label{two-point}
\Pi_{\mu \nu}(p)&=\frac{2}{(2\pi)^4}\int d^4q~tr~(\gamma_\mu\frac{1}{\s q-m}\gamma_5\s b \frac{1}{\s q-m}\gamma_{\nu}\frac{1}{\s q+\s p-m})\nonumber \\&=-\frac{8i\epsilon^{\mu\nu\rho\sigma}}{(2\pi)^4}\int d^4q~\frac{-2q.b q_\rho(q+p)_\sigma+q^2b_\rho(q+p)_\sigma+m^2(q-p)_\rho b_\sigma}{(q^2-m^2)^2((q+p)^2-m^2)}
\end{align}
\end{widetext}
in which the extra factor of two is coming from having two corresponding Feynman diagrams, FIG.~\ref{oneloopvertex}. In \ref{two-point}  the minus sign from fermion loop cancels the minus sign from the chiral-chemical potential vertex.  

As we explained in section \ref{divergences}, the  linearly divergent integrals are the source of ambiguity in Chern-Simons coefficient. The integrals that diverge slower will be independent of our regulator. We use this fact and use our regulator to find only linearly divergent terms.
With our regulator turned on, all the integrals are finite and smooth, meaning that they can be Taylor expanded in terms of external momentum.  From definition of Chern-Simons's coefficient $k_\mu$, the linear part of two-point function \ref{two-point} in external momentum is proportional to this coefficient,
\begin{align}
\Pi^{\mu \nu}(p)=\epsilon^{\mu\nu\rho\sigma}M_{\rho}^{\lambda}b_\lambda p_\sigma~~~s.t.~~~k_\mu=-\frac{1}{2}M_{\mu}^{\lambda}b_\lambda
\end{align}
in which $M_{\rho}^{\lambda}$ is a constant.\\

Let us investigate the most general linearly divergent integral that may be present in our calculations and then come back to the evaluation of \ref{two-point}. The most general linearly divergent integral we face can be written in the following form
\begin{align}
I_{\mu\nu\rho}(p)&= \int d^4q~\frac{q_\mu q_\nu q_\rho}{(q^2-m^2)^2((q+p)^2-m^2)}~~~s.t.~~~\mu\neq\nu\nonumber \\&=
I_{\mu\nu\rho}(0)+\partial_\lambda I_{\mu\nu\rho}(0)~p^\lambda+O(p^2)
\end{align}
By dimensional analysis, we can see that the second term is no longer linearly divergent, as it 
is proportional to $p_\mu$, as a result we only should be concerned about the first term, $I_{\mu\nu\rho}(0)$,
\begin{align}
I_{\mu\nu\rho}(0)&= \int d^4q~\frac{q_\mu q_\nu q_\rho}{(q^2-m(q)^2)^3}~~~~s.t~~~~\mu\neq\nu\
\end{align}

As a first observation, we note that by dimensional analysis $I_{\mu\nu\rho}(0)$ is free of IR divergences and the momentum independent part of $m$ can be set to zero .
At the same time, for the general form of mass-like regulator, like what we found in [\ref{A1}], the linear term would be absent in $m$, as we assume that the low energy theory is already chosen.  Another comment is that the coefficient of the regulator($m$) won't play a role in our calculations, and can be taken care of by a simple renaming of the momentum. Doing so, we find that the divergences are linear in that coefficient. Finally, from the symmetry of integral, we find that for symmetric regulators such that $m(p_\mu)=m(-p_\mu)$ and in particular the regulator we found in [\ref{A1}], the integral vanishes. \\

We saw that  $I_{\mu\nu\rho}(0)$ vanishes for our regulator, we then only need to consider $\partial_\lambda I_{\mu\nu\rho}(0)~p^\lambda$ in order to evaluate  $I_{\mu\nu\rho}(p)$. Let's look at each linearly divergent term in \ref{two-point} separately. The non-vanishing part of the first term in \ref{two-point} has the form of, \[p^\lambda\partial_\lambda \int d^4q~\frac{q.b q_\rho q_\sigma}
{(q^2-m^2)^2((q+p)^2-m(q+p)^2)}|_{p=0}\]
which makes the evaluation of its contribution easy. This integral gets contracted with the Levi-Civita tensor and subsequently vanishes.
The second term has a part that is linearly divergent and is given by,
\begin{align}
p^\lambda&\partial_\lambda \int d^4q~\frac{q^2 q_\sigma}{(q^2-m[q]^2)^2((q+p)^2-m[q+p]^2)}|_{p=0}\nonumber \\&=-2 \int d^4q~\frac{q_\sigma q_\lambda}{(q^2-m^2)^4}q^2p^\lambda~~~~s.t~~~~\rho\neq\sigma\nonumber \\&=-2\int d^4q~\frac{1}{(q^2-m^2)^4}\frac{q^2}{4}q^2 p_\sigma 
\end{align}
Here we have used the fact that the integral is not UV divergent and only has IR divergence. As a result we can safely turn off the regulator and only keep a constant non-zero mass to regulate IR divergences(We carried out the same calculation with the UV regulator turned on and confirmed our result.).  As well, we have used the fact that $<q_0^2>=-<q_i^2>$, $<q_\mu^2>$ defined as $<q_\mu^2>=\int d^4q f(q^2) q_\mu^2$ which can be proven by using Wick rotation. Let $q'_0=i q_0$, after this change of variables the integral gets an extra factor of $-i$ coming from $dq_0$ and the integration contour would rotate by $\frac{\pi}{2}$  counter-clockwise. Using the Euler’s theorem, we can replace the integral with the same integral, integrating over real momentum from minus infinity to infinity. In this new parametrization, the metric is Euclidean and it's easy to compare the integrals. We can now use this fact, or explicitly go back to Minkowskian space and find that $<q_0^2>=-<q_i^2>=\frac{1}{4}<q^2>$.\\
We can simplify our calculation by considering the $q_{\rho}$ part of last term in integral \ref{two-point}. As its contribution to the coefficient of the term linear in external momentum is important for us, we can Taylor expand it in $p_\lambda$,
\begin{align}
p^\lambda&\partial_\lambda \int d^4q~\frac{m^2 q_\rho}{(q^2-m[q]^2)^2((q+p)^2-m[q+p]^2)}|_{p=0}  \nonumber\\&=-2 \int d^4q~\frac{q_\rho q_\lambda}{(q^2-m^2)^4}p^\lambda~~~~s.t~~~~\rho\neq\sigma\nonumber \\&=-2\int d^4q~\frac{1}{(q^2-m^2)^4}\frac{q^2}{4} m^2 p_\rho 
\end{align}

We now combine all of the results to find the two-point function. The two-point function simplifies to,
\begin{align}
\Pi^{\mu \nu}(p)=&-\frac{8i\epsilon^{\mu\nu\rho\sigma}}{(2\pi)^4}\int d^4q~\frac{1}{(q^2-m^2)^3}~(-2q.b q_\rho p_\sigma \nonumber \\&~ +q^2b_\rho p_\sigma-m^2 p_\rho b_\sigma-\frac{1}{2}q^2p_\sigma b_\rho)\nonumber \\=&
-\frac{8i\epsilon^{\mu\nu\rho\sigma}}{(2\pi)^4}\int d^4q~\frac{1}{(q^2-m^2)^3}~(-\frac{1}{2}q^2 b_\rho p_\sigma \nonumber \\&~ +q^2b_\rho p_\sigma-m^2 p_\rho b_\sigma-\frac{1}{2}q^2p_\sigma b_\rho)
\nonumber \\=&
-\frac{8i\epsilon^{\mu\nu\rho\sigma}}{(2\pi)^4}\int d^4q~\frac{1}{(q^2-m^2)^3}~\left(-m^2 p_\rho b_\sigma\right)
\end{align}
in second equality we have assumed that $m(p)=m$, is a constant, as the integrals are not linearly divergent.
The remaining integral can be evaluated straightforwardly. It is given by,
\begin{align}
&\int d^4q~\frac{1}{(q^2-m^2+i\xi)^3}=\frac{i\pi^2}{2m^2}
\end{align}
in which $\xi$ is the Feynman regulator, regulating IR divergences. \\
We finally find that the two-point function is given by,

\begin{align}
\Pi^{\mu \nu}(p)=
\frac{\epsilon^{\mu\nu\rho\sigma}}{4\pi^2}b_\rho p_\sigma
\end{align}
then the Chern-Simons coefficient can be extracted,

\begin{align}
 k_\mu=-\frac{1}{8\pi^2}b_\mu
\end{align}
As we mentioned in last sections, we find that the space-like Chern-Simons terms can indeed be present even after regulating by a regulator that breaks emergent Lorentz symmetry manifestly.  The Chern-Simons coefficient that we find is the same as the coefficient found by Perez  \cite{Perez}.

\section{Conclusion}\
In this work we studied in detail 3D Weyl semi-metals in presence of perturbation that break $P$, chiral-chemical potential. 
We argued that there can be two kinds of linear responses associated to a 3D Weyl semi-metal with broken $\mathcal{PT}$. We found that in the meta-stable scenario, there exist a Chern-Simons term in effective action of gauge field which contributes to electric conductivity. We concluded that in equilibrium there can not be any space-like Chern-Simons term.
As a result, in the equilibrium case chiral-magnetic effect will be absent and the conductivity would be gain the  usual Hall conductivity form. 
This effect still remains to be confirmed experimentally.

\section{Acknowledgement}
We thank M.~Franz for suggesting the problem and G.W.~Semenoff, I.~ Affleck, M.~Franz and A.~Zhitnitsky for useful comments.
\section{APPENDIX} 
\appendix

\section{EFFECTIVE HAMILTONIANIAN}\label{A1}
\
In this appendix, we argue that effective Hamiltonian of a band theory without interaction between quasi-particles is given by the low energy limit of the theory. We then rewrite the low energy limit of $H_0$ and $H_1$ in relativistic notations. \\

 A general band theory can be written in terms of a quadratic Hamiltonian given by $\mathcal{H}=\sum_{\alpha, k} \epsilon_\alpha(k) c^\dagger_k c_k$, in which $\alpha$ indexes the bands. As a consequence of  being Gaussian, path-integral formalism can be used to show that high-energy modes can be integrated out and we are left with the modes of interest. The effective theory then is given by the low-energy limit of the Hamiltonian.

Let us find the low-energy limit of our model.  From equation \ref{H_0}, the expansion of $H_0$ around $\bf k=0$ gets the following form 
\begin{align}
H_0=\lambda
\left(\scalemath{0.85}{
\begin{array}{cccc}
0&2 k_y+2i k_x&-2i k_z+\frac{t}{\lambda}k^2&0\\
2 k_y-2ik_x&0&0&-2i k_z+\frac{t}{\lambda} k^2\\
2i k_z+\frac{t}{\lambda} k^2&0&0&-2 k_y-2i k_x\\
0&2i k_z+\frac{t}{\lambda} k^2&-2 k_y+2ik_x&0 
\end{array}}
\right)
\end{align}
From the definition of the lowest order term in $H_0$, in equation \ref{H_02}, the $\alpha$ matrices can be extracted. For example for $\alpha_1$, we find that, 
\begin{align}
\alpha_1=-\sigma_zs_y=\left(\begin{array}{cccc}
0&i&0&0\\
-i&0&0&0\\
0&0&0&-i\\
0&0&i&0 
\end{array}
\right)
\end{align}
The rest of the matrices are $\alpha_2=\sigma_zs_x,~\alpha_3=\sigma_y1_{2\times 2},~\beta=\sigma_x1_{2\times 2}$.
To extract $\beta$, which plays the role of a mass term, $H_0$ must be expanded in the region that the quasi-particles are massive($ \epsilon\neq 6t$).

The Dirac $\gamma$ matrices are as usual defined by, 
\begin{align}
&\gamma_i=\beta \alpha_i \nonumber \\ &
\gamma_0=\beta
\end{align}
and $\gamma_5$ in 3D can be defined by,  
\begin{align}
\gamma_5&=i\gamma_0\gamma_1\gamma_2\gamma_3\nonumber\\ &=\sigma_ys_z
\end{align}
It can be checked that $\gamma$ matrices satisfy Clifford Algebra defined by $\{\gamma_\mu,\gamma_\nu\}=2\eta_{\mu\nu}$, $\eta$ being the metric, $\eta=\text{diag}(1,-1)$. Now that we have an explicit form for $\alpha$ matrices, we can write $H_1$ and the second order part of $H_0$ in terms of these matrices. They are given by
\begin{align}
H_1(k)&=b_0 \gamma_5+{\bf \a.b}  \gamma_5
\nonumber \\&=b_0 \gamma_5+b_i \gamma_0 \gamma_i \gamma_5
\end{align}
\begin{align}
H_{\text{regulator}}(k)&=\frac{t}{\lambda} {\bf k}^2 \beta\nonumber \\
&=\frac{t}{\lambda} {\bf k}^2 \gamma_0
\end{align}

\end{document}